# A vulnerability in Google AdSense: Automatic extraction of links to ads

Prof. Ph. D. Manuel Blázquez
Universidad Complutense de Madrid
manublaz@ucm.es

**ABSTRACT**

On the basis of the XSS (Cross Site Scripting) and Web Crawler techniques it is possible to go through the barriers of the Google Adsense advertising system by obtaining the validated links of the ads published on a website. Such method involves obtaining the source code built for the Google java applet for publishing and handling ads and for the final link retrieval. Once the links of the ads have been obtained, you can use the user sessions visiting other websites to load such links, in the background, by a simple re-direction, through a hidden iframe, so that the IP addresses clicking are different in each case.

## 1. INTRODUCTION

The correct encryption and coding of the ads published by third parties is a problem of paramount importance when it comes to ensure the confidence and reliability of the pay per click systems. With this method, the advertising company pays the websites publishing their advertisements, provided that the users click on them, in a proportionate and transparent manner depending on the number of visits and their validation. This system is used by Google Adsense, Google's advertising service, which manages the suitable ads for each website and user profile. When it comes to studying Google's advertising system the following questions arise .What's the security level of the ads? Can you obtain links automatically? Is it possible to click automatically on the ads obtained? In the paper by (Mann, C.C. 2006) published in Wired magazine, they lend importance to the fact that there is a prevalence of mistrust due to click frauding in advertising systems such as Google AdSense. Specifically it is argued that there are robots capable to boost the number of clicks of an ad published on a website, fostering the profit of the advertising platform.

To give an answer to those questions a security test was developed. Such test combined Cross Site Scripting (XSS) and Web Crawler techniques to overcome the system barriers to get the legitimate and validated links generated by Google Adsense during the user's visit to the website. After a positive result, in October 2013 a first round of talks was held to warn Google about the security breach detected and to work together patching the loophole. In January 2014 Google neither had admitted nor denied the existence of such problem, thus to this day it has not been solved. In the present paper we want to draw the attention about this kind of vulnerabilities that may be resolved and ought to be taken into account because of the danger they entail.

## 2. METHODOLOGY

Google Adsense vulnerability can be sequenced in two phases. 1) Obtaining the links of the original ads of a website automatically. 2) Execute automatic clicks on the ads obtained.

**1) Obtain the links of the original ads of a website automatically**

The vulnerability target is all the websites which contain Google Adsense text ads. In any case, the links of the ads are not visible by observing the website source code. However, it is possible to check the URL address of the link by analyzing the ad web element with a *DOM browser*. Thus, there are security mechanisms in Google AdSense which enable to build the HTML elements of the ads in DOM and java, once the website has been loaded in the client's browser. It is for this reason that the links of the ads are not visible in the web source code despite being there. This security measure is designed as to prevent the links from being obtained automatically and it is deemed to be an unassailable technique (GANDHI, M.; JAKOBSSON, M.; RATKIEWICZ, J. 2006, p.134). Unfortunately, it is possible to breach this barrier; you only have to obtain the target website source code and detect the Google AdSense code, see figure 1.

```
<div id="GoogleAdSense">
<script language="JavaScript" type="text/javascript">
        google_ad_client = "pub-1229649499684927";
        google_ad_channel = "ANUNCIOS.COM"
        google_ad_type = "text";
        google_max_num_ads = 3;
        google_language = "es";
        google_safe  = "high";
        google_encoding = "utf8";
        google_ad_width = 336;
        google_ad_height = 280;
        google_ad_format = "336x280_as";
        google_color_border = "EEEEEE";
        google_color_bg = "EEEEEE";
        google_color_link = "000066";
        google_color_text = "000000";
        google_color_url = "CC0000";
</script>
<div id="GoogleAd">
<span>Sponsored links </span>
<script language="JavaScript" src="http://pagead2.googlesyndication.com/pagead/show_ads.js"
type="text/javascript"></script>
</div>
</div>
```
*Figure 1. Java code in Google AdSense*

The java file "show_ads.js" embeds the ads in the target website HTML code once it has been completely loaded in the browser. Nevertheless, they are not introduced directly, they use two window frames <iframe> which we will call "Iframe 1" and "Iframe 2", see figure 2.

```
<!-- Iframe 1 -->
<iframe width="0" height="0" frameborder="0" marginwidth="0" marginheight="0" vspace="0" hspace="0"
allowtransparency="true" scrolling="no" allowfullscreen="true" style="display:none" id="google_esf"
name="google_esf"
src="http://googleads.g.doubleclick.net/pagead/html/r20140904/r20140417/zrt_lookup.html"></iframe>

<!-- Iframe 2 -->
<iframe id="google_ads_frame1" name="google_ads_frame1" width="336" height="280" frameborder="0"
src="http://googleads.g.doubleclick.net/pagead/ads?client=ca-pub-
1229649499684927&format=336x280_as&output=html&h=280&adk=3476478377&w=336&lmt=1410
413197&num_ads=3&channel=ANUNCIOS.COM&ad_type=text&color_bg=EEEEEE&color_border=EEEEEE
&color_link=000066&color_text=000000&color_url=CC0000&oe=utf8&flash=14.0.0&hl=es&a
mp;url=http%3A%2F%2Flocalhost%2Fvigilante%2Fexploits%2Fexploit2.php&adsafe=high&dt=1410420397140&a
mp;bpp=10&bdt=23&shv=r20140904&cbv=r20140417&saldr=sa&correlator=8467094044672&frm
=20&ga_vid=482323067.1410420397&ga_sid=1410420397&ga_hid=2126412725&ga_fc=0&u_tz=120&a
mp;u_his=2&u_java=1&u_h=900&u_w=1600&u_ah=856&u_aw=1600&u_cd=24&u_nplug=14&
;u_nmime=100&dff=times%20new%20roman&dfs=16&adx=8&ady=8&biw=1600&bih=795&eid=3
17150304%2C317150313&oid=3&rx=0&eae=0&fc=8&brdim=0%2C0%2C0%2C1600%2C0%2C1600%2C856
%2C1600%2C795&vis=1&abl=CS&ppjl=u&fu=0&ifi=1&xpc=EvP9CmkS4y&p=http%3A//localho
st&dtd=50" marginwidth="0" marginheight="0" vspace="0" hspace="0" allowtransparency="true"
scrolling="no" allowfullscreen="true"></iframe>
```
*Figure 2. Insertion of the show_ads.js iframes*

Iframe 1 contains a Google AdSense code cross-checking and integrity verification method within the website being loaded (Linden, J.; Teeter, T. 2006). This system allocates the ad identifier and user necessary to execute the code inserted in Iframe 2, hampering the handling of the ad source code for automatic extraction. Furthermore, Iframe 2 loads a dynamic website containing the ads themselves.

Thus, to make a valid loading of the ads in Iframe 2, permitted by Iframe 1, it is necessary to execute all the Google AdSense code and subsequently extract the link of the Iframe 2 dynamic website. To do so you can use XSS and JavaScript techniques. In particular a form called "technical1", with a field storing the Iframe 2 URL address, is added to the "GoogleAdSense" layer. This is achieved through the instructions shown in figure 3.

Method of replacing the target website source code. We can see that the GoogleAdSense layer is written again with the form "technical1" that will be used as data recipient.

```
$html1 = preg_replace("/<div id=\"GoogleAdSense\">/", "<div id=\"GoogleAdSense\"> + Form technical1", $html1);
```

Form "technical1", introduced within the "GoogleAdSense" layer containing the "code1" text area, where the "Iframe2" URL address is to be stored once it is executed in the website.

```
<div id="GoogleAdSense">
    <div>
      <form name='technical1' action='$_SERVER[PHP_SELF]' method='post'>
      <textarea id='code1' name='code1'></textarea>
      </form>
    </div>
</div>
```

JavaScript code to extract and copy the "Iframe2" URL address and to paste it in the "code1" text area of the "technical1" form.

```
<script>
window.onload = init;
function init() {
document.getElementById('code1').value=document.getElementById('google_ads_frame1').src;
document.technical1.submit();
}
</script>
```

*Figure 3. Code to extract the link of Iframe 2*

Figure 3 script is executed once the website is loaded, like the AdSense code, but subsequently to allow Iframe 1 and 2 to be loaded. Then a value will be assigned to the field of "code1" form corresponding to the Iframe 2 URL address, identified as "google_ads_frame1". Once obtained, "technical1" form will be transmitted with such information. In figure 2 it is noticeable that the Iframe 2 URL address is not correct, as it introduces among its variables the domain address from which the code is being executed: see URL "http://localhost/vigilante/exploits/exploit2.php". By clicking on the ad without modifying these values of related pages, Google AdSense could detect some sort of incoherence. To avoid it Iframe 2 URL address shall be modified in the **&url** and **&p** variables, see figure 4.

```
$code = $_POST[code1];
$code = preg_replace("/\&url=.*\&adsafe/", "&url=http://www.anuncios.com/&adsafe", $code);
$code = preg_replace("/\&p=http.*/", "&p=http://www.anuncios.com", $code);
```
*Figure 4. Modification of the link of Iframe 2*

When the link of Iframe 2 is retrieved, the variables **&url** and **&p** will be replaced by the domain address of the studied website, in this case *http://www.anuncios.com*, through regular expressions. Once the URL address of Iframe 2 is prepared, the source code is retrieved, with the function "file_get_contents", directly accessing to the links of the ads already validated by Google AdSense. During this process, you can use DOM to load the HTML structure of the page and XPath to obtain only the available links, see figure 5.

```
$html2 = file_get_contents("$code");
$dom2 = new DOMDocument();
@$dom2->loadHTML($html2);
$xpath2 = new DOMXPath($dom2);
$links = $xpath2->query("/html/body//a");
```
*Figure 5. Obtaining the source code of the Iframe 2 page and retrieval of the links of the ads*

The result of applying this method is the links of the ads, as shown in figure 6. To test the functioning, this program can be downloaded at: http://www.mblazquez.es/docs/google-ads-extractor.zip. You can also see how it works at: https://youtu.be/0tIBcJ-VN7s

```
http://googleads.g.doubleclick.net/aclk?sa=l&ai=CCWOcql4RVJDHLIHCiQa1zoCICa-
GqZsH75Pj9aIBwI23ARABIPmY_AEoA1DA6qP9-
v____8BYNW11wKgAdHQoN4DyAEBqQKMWfG0Fae1PqgDAcgDwwSqBG5P0Ig3k_RIM7eChCXVjUsXttfDLlj6od6JU7FeWhdmg0GwM8-
w6Nhuh9awpyuPhzyt-gKK2kLj6_Fp03lAl8FoHzX6xTk8nztGi79o9q9viXDkav02-Yi2F5OT67hR-
ItZO2J04gvpcTvvajkQpYAHl6_fIQ&num=1&sig=AOD64_2umGsoLzRoV9h4YazaUGzHORn0xQ&client=ca-pub-
1229649499684927&adurl=http://candidatos.sanroman.com/resultado-
busqueda.php%3Ffiltro%3D%26area%3DMarketing%26sector%3D0%26modalidad%3D%26provincia%3DMADRID%26seccion%3D

http://googleads.g.doubleclick.net/aclk?sa=l&ai=Cj0h3ql4RVJDHLIHCiQa1zoCICfGSgIoF2Z2bkJcBwI23ARACIPmY_AEoA
1DwkovEBGDVtdcCoAH___T8A8gBAagDAcgDwwSqBG5P0NgMkPRLM7eChCXVjUsXttfDLlj6od6JU7FeWhdmg0GwM8-
w6Nhuh9awpyuPhzyt-gKK2kLj6_Fp03lAl8FoHzX6xTk8nztG1dsY_a9viXDkav02-Yi2F5OT67hR-ItZO2J04gvpcTvvEC0ehYAH6f-
KAw&num=2&sig=AOD64_2PS3pUS7ccIZ8gg9nxsoRo1dgCCg&client=ca-pub-
1229649499684927&adurl=http://es.emailbrain.com/ebs/index.shtml%3FMedium%3DPPC%26Campaign%3DES_Email_Marke
ting%26Adgroup%3DEmail_Marketing_-_Broad

http://googleads.g.doubleclick.net/aclk?sa=L&ai=C4_Hlql4RVJDHLIHCiQa1zoCICamd6N0E4def75EBwI23ARADIPmY_AEoA
1CJtf3UB2DVtdcCyAEBqQKMWfG0Fae1PqgDAcgDwwSqBGtP0Khal_RKM7eChCXVjUsXttfDLlj6od6JU7FeWhdmg0GwM8-
w6Nhuh9awpyuPhzyt-gKK2kLj6_Fp03lAl8FoHzX6xXE8Vwfifvy9qq9vw3DkOuA2-
n2112aQK0xS6H9aK9d34qvv2A9jrYAH2Zf5Lg&num=3&sig=AOD64_1aYwxLbhcZISbWwRtDQhgLRfddmg&client=ca-pub-
1229649499684927&adurl=http://www.banderasysoportes.com
```
*Figure 6. Links to ads obtained automatically*

The method of automatic extraction of links of AdSense ads can be summarized in the steps set forth below: a) extraction of the target website source code, b) detection of the Google AdSense Java code, c) Execution of the Google AdSense Java code, d) Iframe 1 and 2 loading, e) Insertion of the technical1 form and code1 field, f) Extraction of Iframe2 URL address to code1 field, g) Submission of the form to the same execution environment, h) Preparation of the URL of Iframe2 transmitted in the form, i) Obtaining the URL address source code of iframe 2, j) Extraction of links of the ads.

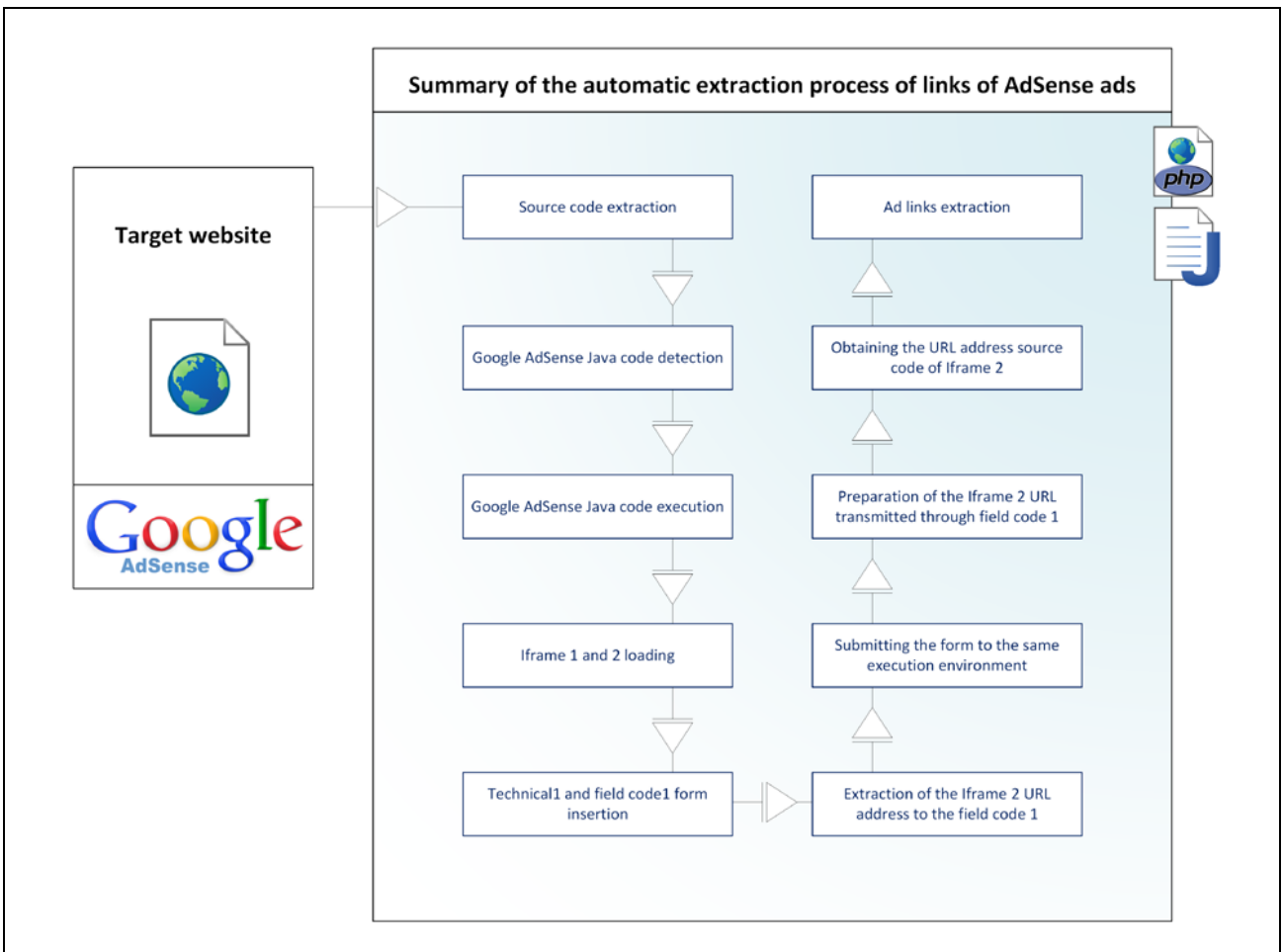

*Figure 7. Summary of the automatic extraction process of links of AdSense ads*

**2) Executing automatic clicks on the ads obtained**

After demonstrating the capacity to extract the links of the ads automatically, it doesn't sound strange to click on the links of the ads obtained. Such task may be complex because Google AdSense has been putting in place statistical security measures able to detect fraudulent clicks. This filter system (Google Support. 2014) is hypothetically capable to recognize the clicks whose objective is to increase the advertising expenses to raise the profits of the website owners hosting their ads. In such a case they are automatically disabled. According to (Adwords Blogspot. 2007) around 10% of the registered clicks are fraudulent and detected automatically by the system, in 0,02% of the cases a specialized research is necessary to determine their validity. Likewise it is also to be noted that it you can have false positives in the process, as stated by (Dave, V.; Guha, S.; Zhang, Y. 2012, p.178) in his paper about click-spam quantification in advertisement webs.

Going further into the topic of fraudulent clicks detection, according to (Kshetri, N. 2010), the current security systems may be classified in: a) Anomalies detection, b) Heuristic detection of invalid clicks depending on several variables such as the proportion, clicks ratio, browsing path. c) Through a classification of users' patterns of behavior and clicks. For Google AdSense, it is very likely that they incorporate a combination of such methods on the basis of published patents on the subject.

For example, the identification method for advertisement processing (Li, Z.; Ou, C.; Park, S.U.; Savoor, R.; Sposato, S. 2007) designed to test the IP and the time spent by the user on the website where he clicks. If we are facing an automatic click, the time spent by the user would be close to zero or the clicking IP would be the same.

It is also to mention the patent for the detection of fraudulent clicks of (Gillespie, J.; Meggs, A.F. 2007) which records each user click and the research and browsing path inputting information such as the clicks per minute rate, the click coverage ratio, the average number of visits and clicks on the website, the user visits and clicks ratio, among other items. This method assumes that a deviation above normal, with regard to the increase of the visits and clicks statistics in a website, involves the automatic detection of fictitious clicks. The weakness of this method lies in the possibility of disguising statistically the clicks, increasing the number of visits and controlling the time spent on the websites, before executing the automatic click.

The patent for the detection of fraudulent clicks of (Zwicky, R.K. 2010) puts forward a stepped system that takes into account the advertiser for fraud detection, providing supervision panels to report suspicious situations, like for example IP addresses from visitors considered to be robots, sudden click increase and aberrations in the browsing of the user clicking.

On the methods for fraudulent click detection through automatic classification, (Yan, J. H.; Jiang, W. R. 2014) patent is utterly essential. It is a system which registers with tabs the users accessing the target website, identifying their browsing and clicks on advertisements. Thanks to this method, when the patterns do not coincide with the ordinary use of the website, or with the usual number of visits, the user will be considered a robot and his clicks turn out to be fraudulent. This fact is registered and serves as experience for the detection of further malicious clicks.

Whereas there are many security measures, their complete vulnerability appears to be a process difficult to determine without Google's collaboration. Despite this fact, security problems may arise if the following premises were met:

a) **Websites group for user attraction.** One way to avoid user IP addresses to be always the same is to use a Websites group with an uninterrupted and varied worldwide number of visits. Each visit may be used to load in an iframe window the program necessary to extract the links of the ads of a target website and click on them. Therefore the time factor and the user IP address would be resolvable. Still, this method is not foolproof, because Google can notice that the number of visits of the target website which has the advertisements is not customary and thus register more clicks than the initially likely to occur. In that case it is possible to redirect visits without clicking on the ads, in order to increase the number of visits of the site and thereby the traffic, in a variable and progressive manner. If the traffic rises, the possibility to click on the website ads also grows proportionately.

b) **Program for the redirection, link extraction and automatic click.** A JavaScript program can generate an invisible Iframe window on the websites aimed to attract users, this will allow to load at the same time the target website containing the ads to be clicked and will execute the link extraction method previously proposed to finally decide, depending on the statistics if you have to click or not on the advertisement link or if you have to wait. This system could violate the heuristic methods, patterns and automatic classification, while acquiring the normal users of the website own habits.

These security problems become a violation strategy to make automatic clicks, see figure 8.

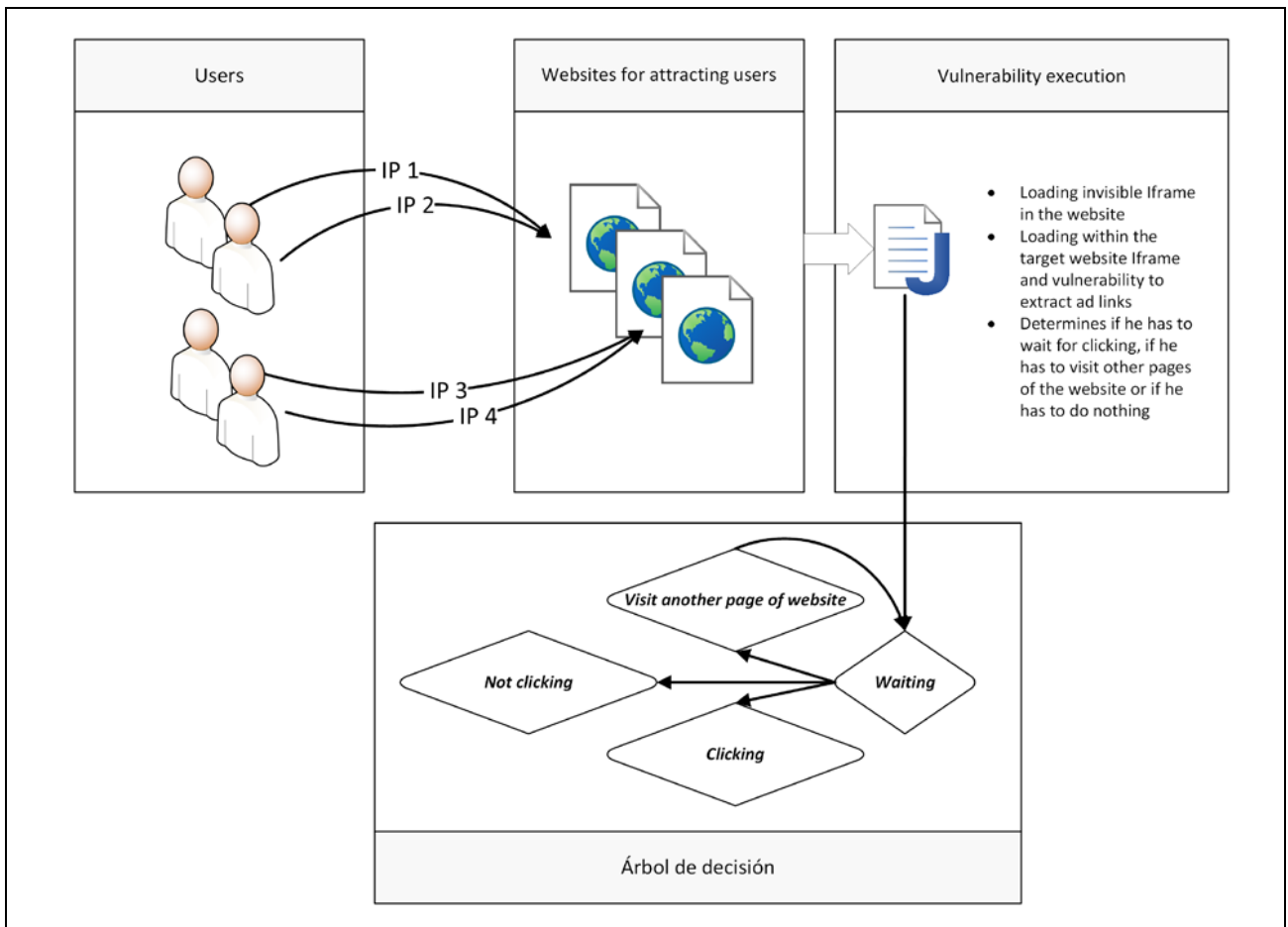

*Figure 8. Automatic click strategy*

The users accessing from several IP addresses, visit websites which "seize" their user session through the execution of a Java code loading an invisible Iframe, in the page they are visiting. However, in the invisible iframe the target website with the ads is loaded and the automatic ad links extraction program is executed at the same time. Once stored, the program can be designed to take several decisions, for example; a) Wait for a random or statistically defined period, b) Visit another page of the website to simulate another query from the user, c) Not to click depending on the website ratio, d) Click on an ad. The vulnerability is executed as many times the user accesses to the capturing websites, obtaining for each of them different advertisements, customized for each profile, as Google AdSense would do it without vulnerability.

To test the automatic click strategy hypothesis, a program which simulates a user capturing blog, called "blog1" and a target blog which contains ads, called "blog2" has been developed. Blog1 executes a "google_analytics_top.js" java code allowing the introduction within the layer with "*booster*" identifier of two other layers called "sub1" and "sub2" as well as a new iframe containing the code for the extraction of the links of the ads in blog2 and the loading process of the later, using programmed rules and patterns simulating a normal behavior of the user. Once you click on the advertisement, the program erases any trace removing any reference to the link contained in the layer. This program can be downloaded at:
http://www.mblazquez.es/docs/booster.zip.

## 3. CONCLUSIONS

Google AdSense dazzles the ad codes, hindering the access to his links, to avoid them to be used for malicious purposes. To that end they avoid viewing their source code through the two embedded iframes and generated with java. This way, the specialized robots cannot retrieve automatically the links of the ads.

However, the security of the published ads in the websites associated to the Google Adsense program is potentially compromised because it is possible to retrieve automatically his links with the Google validation code, as it has been proven. This technique may be used to raise the profits in the pay per click system or to disable the ad publishing platform accounts from the competitors if the statistical limits are surpassed.

Google AdSense does not protect so well the links of the ads as the detection systems afterwards. This means that the fictitious and true clicks must be filtered later. In fact Google uses up to 5 patents specialized in the statistical detection of fraudulent clicks.

Taking into account that the detection systems by fraudulent click are based in the normal browsing patterns and user interaction, they could be overcomed applying a user capturing and re-direction method, combined with the advertisement automatic extraction. It is possible to simulate the user browsing in a website and click in a simple or coordinated way with the wished statistics.